\documentclass[useAMS,usegraphicx]{mn2e}
\usepackage{color}

\usepackage{times}
\newcommand{\sqcm}{cm$^{-2}$}
\newcommand{\fek}{Fe~K$\alpha$}
\newcommand{\xmm}{{\em XMM-Newton}}
\newcommand{\nustar}{{\em NuSTAR}}
\newcommand{\chandra}{{\em Chandra}}
\newcommand{\swift}{{\em Swift}}
\newcommand{\suzaku}{{\em Suzaku}}
\newcommand{\sax}{{\em BeppoSAX}}
\usepackage[space]{grffile}
\usepackage[]{amsmath}
\usepackage{amssymb}
\usepackage[normalem]{ulem}
\usepackage{natbib}
\usepackage{hyperref}
\bibpunct{(}{)}{;}{a}{}{,}
\usepackage{comment}

\begin{document}

\title[The NuSTAR X-ray spectrum of NGC~7213]{The NuSTAR X-ray spectrum of the low-luminosity AGN in NGC~7213}
\author[F. Ursini et al.]
{F. Ursini,$^{1,2,3}$\thanks{e-mail: \href{mailto:francesco.ursini@obs.ujf-grenoble.fr}{\texttt{francesco.ursini@obs.ujf-grenoble.fr}}}
A. Marinucci,$^{3}$
G. Matt,$^3$
S. Bianchi,$^3$
A. Tortosa,$^3$
D. Stern,$^4$
\newauthor
P. Ar\'evalo,$^{5}$
D.~R. Ballantyne,$^6$
F.~E. Bauer,$^{7,8,9}$
A.~C. Fabian,$^{10}$
\newauthor
F.~A. Harrison,$^{11}$
A.~M. Lohfink,$^{10}$
C.~S. Reynolds,$^{12,13}$
D.~J. Walton$^{14,4}$\\
$^1$ Univ. Grenoble Alpes, IPAG, F-38000 Grenoble, France. \label{ipag}\\
$^2$ CNRS, IPAG, F-38000 Grenoble, France. \label{cnrs}\\	
$^3$ Dipartimento di Matematica e Fisica, Universit\`a degli Studi Roma Tre, via della Vasca Navale 84, 00146 Roma, Italy. \label{rome}\\
$^4$ Jet Propulsion Laboratory, California Institute of Technology, 4800 Oak Grove Drive, Mail Stop 169-221, Pasadena, CA 91109, USA. \label{jpl}\\
$^5$ Instituto de F\'isica y Astronom\'ia, Facultad de Ciencias, Universidad de Valpara\'iso, Gran Bretana Nº 1111, Playa Ancha, Valpara\'iso, Chile.\\
$^6$ Center for Relativistic Astrophysics, School of Physics, Georgia Institute of Technology, Atlanta, GA 30332, USA. \label{git}\\
$^7$ Instituto de Astrof\'isica, Facultad de F\'isica, Pontificia Universidad Cat\'olica de Chile, Casilla 306, Santiago 22, Chile.\label{santiago}\\
$^8$ Millennium Institute of Astrophysics, Vicu\~{n}a Mackenna 4860, 7820436 Macul, Santiago, Chile.\\
$^9$ Space Science Institute, 4750 Walnut Street, Suite 205, Boulder, CO 80301, USA.\label{ssi}\\
$^{10}$ Institute of Astronomy, University of Cambridge, Madingley Road, Cambridge CB3 0HA. \label{cambridge}\\ 
$^{11}$ Cahill Center for Astronomy and Astrophysics, California Institute of Technology, Pasadena, CA 91125, USA.\label{caltech}\\
$^{12}$ Department of Astronomy, University of Maryland, College Park, MD 20742-2421, USA.\\
$^{13}$ Joint Space-Science Institute (JSI), College Park, MD 20742- 2421, USA.\\
$^{14}$ Cahill Center for Astronomy and Astrophysics, California Institute of Technology, Pasadena, CA 91125, USA.
}

\date{Released Xxxx Xxxxx XX}


\maketitle

\label{firstpage}

\begin{abstract}
We present an analysis of the 3--79 keV \textit{NuSTAR} spectrum of the low-luminosity active galactic nucleus NGC~7213. In agreement with past observations, we find a lower limit to the high-energy cut-off of $E_c > 140$ keV, no evidence for a Compton-reflected continuum, and the presence of an iron K$\alpha$ complex, possibly produced in the broad-line region. From the application of the \textsc{MYTorus} model, we find that the line-emitting material is consistent with the absence of a significant Compton reflection if arising from a Compton-thin torus of gas with a column density of $5.0^{+2.0}_{-1.6}\times 10^{23}$ \sqcm. We report variability of the equivalent width of the iron lines on the time-scale of years using archival observations from \textit{XMM-Newton}, \textit{Chandra} and \textit{Suzaku}. This analysis suggests a possible contribution from dusty gas. A fit with a Comptonization model indicates the presence of a hot corona with a temperature $kT_e >40$ keV and an optical depth $\tau \lesssim 1$, assuming a spherical geometry.
\end{abstract}

\section{Introduction}
The central engine of low-luminosity active galactic nuclei (LLAGNs) is thought to be powered by accretion of surrounding matter on to a supermassive black hole, similar to more powerful AGNs, like Seyfert galaxies and quasars \citep[see][for a review]{ho2008review}. The X-ray spectrum of AGNs is generally dominated by a primary power law component, which is thought to be produced by Comptonization of optical/UV photons emitted by the underlying accretion disc in a hot plasma, the so-called corona \citep[see, e.g.,][]{haardt&maraschi1991,hmg1994,hmg1997}. A signature of this process is the presence of a high-energy cut-off in the X-ray emission, which has been observed in a number of sources \citep[see, e.g.,][]{perola2002,2041-8205-782-2-L25,IC4329A_Brenneman,marinucci2014swift,ballantyne2014,balokovic2015,5548}.\\ \\
The distinctive characteristic of LLAGNs is their intrinsic faintness ($L_{\textrm{bol}} < 10^{44}$ erg~s$^{-1}$). Moreover, the mass accretion rate of LLAGNs is generally small; in terms of the Eddington ratio, most of them have $L/L_{\textrm{Edd}} < 10^{-2}$ while luminous AGNs have $L/L_{\textrm{Edd}} \sim 0.01 - 1$ \citep[see, e.g.,][]{panessa2006,kollmeier2006}. Whether LLAGNS are simply a scaled-down version of classical AGNs is a matter of debate. A standard geometrically thin, optically thick accretion disc \citep{ss1973} may power LLAGNs \citep{maoz2007}, as is commonly assumed for luminous AGNs. However, radiative-inefficient accretion flows \citep[RIAFs; see, e.g.,][]{adaf} have been proposed to explain some observational properties, in particular their lack of a UV bump \citep[see, e.g.,][]{ho2009,yu2011}. Furthermore, in luminous AGNs the hard X-ray photon index and the Eddington ratio are positively correlated \citep[see, e.g.,][]{sobolewska&papadakis}, while an anticorrelation is found in LLAGNs  \citep{gu&cao2009}. This result is consistent with the X-ray emission of LLAGNs originating from Comptonization in RIAFs, and it suggests a similarity between LLAGNs and black-hole X-ray binaries (BHXRBs) in the low/hard state, where the accretion rate is low \citep{wu&gu2008}. The low accretion rate might not only affect the structure of the inner accretion flow, but also that of the putative obscuring torus. The torus is predicted to disappear in low-luminosity sources by models depicting it as a clumpy wind, arising from the outer accretion disc \citep[see, e.g.,][and references therein]{elitzur2006}, or as a collection of many self-gravitating, dusty molecular clouds \citep[see, e.g.,][]{honig2007}.\\ \\
NGC~7213 is a nearby \citep[$z=0.005839$, as given in the NASA/IPAC Extragalactic Database; distance 25.80 Mpc,][]{D7213} LLAGN with $L_{\textrm{bol}} = 1.7 \times 10^{43}$ erg~s$^{-1}$ \citep{emma2012} that hosts a supermassive black hole of $\sim 10^8$ solar masses \citep[estimated from the stellar velocity dispersion, see][]{woo&urry2002}, yielding an Eddington ratio of $1.4 \times 10^{-3}$. It has been historically classified as a Seyfert 1 because its optical spectrum shows broad emission lines, i.e. with a full width at half maximum (FWHM) of a few thousand km/s \citep{phillips1979}. However, it has also been classified as a low-ionization nuclear emission region galaxy (LINER) because of the low excitation observed in the narrow-line spectrum \citep[][]{filippenko1984}. Low-ionization lines were also detected in the soft X-ray band \citep{starling2005}. \citet{wu1983} measured a UV flux higher than the extrapolated optical flux, thus indicating a possible UV bump, but it is still weaker than in most Seyfert galaxies. More recently, \citet{starling2005} found no evidence for an optical/UV bump using \xmm/OM data.\\ \\
The X-ray spectrum of NGC~7213 shows peculiarities as well. The Compton reflection component is found to be weak or absent with \xmm, \sax~and \suzaku~observations \citep[see, e.g.,][]{bianchi20037213,lobban2010}, in contrast to what is commonly observed in Seyfert 1 galaxies. This may suggest that the accretion disc is truncated in the inner region, perhaps replaced by a Compton-thin RIAF \citep{lobban2010}. An iron line complex is clearly detected between 6.4 and 7 keV, consisting of three narrow K$\alpha$ emission lines respectively from neutral Fe, Fe \textsc{xxv} and Fe \textsc{xxvi} \citep[see, e.g., the analysis of \chandra~data by][]{bianchi72132008}. Given the lack of an observed Compton reflection hump, such lines cannot originate from Compton-thick material, like the accretion disc or a parsec-scale torus. \citet{bianchi72132008} found the FWHM of the neutral Fe K$\alpha$ line to be consistent with that of the H$\alpha$ line ($\sim 2600$ km/s), thus suggesting a common origin in the broad-line region (BLR). Another possibility is reprocessed emission from dusty gas, which can produce strong, neutral \fek~emission with a weak associated continuum \citep{gohil2015}. \citet{emma2012} found an anticorrelation between the X-ray photon index and the X-ray luminosity that, together with the low accretion rate of the source, indicates a spectral behaviour similar to that of BHXRBs in the hard state. Finally, \citet{bianchi20037213} reported a high-energy cut-off of $95^{+75}_{-45}$ keV using \textit{XMM-Newton}+\sax~data, while \citet{lobban2010} found a lower limit of 350 keV using \textit{Suzaku}/XIS+PIN and \textit{Swift}/BAT data.\\ \\
In this paper, we report on a \textit{NuSTAR} observation of NGC~7213 performed in October 2014. The primary focus of this work is modelling the broad-band X-ray spectrum of the source and constraining the origin of the X-ray emission. In Sect. \ref{sec:obs&data}, we describe the observations and data reduction. In Sect. \ref{sec:analysis} we present the analysis of the 3--79 keV spectrum, fitted with both a phenomenological and a realistic Comptonization model, and we study the time evolution of the iron line complex by analysing archival \xmm, \chandra~and \suzaku~data. In Sect. \ref{sec:discussion}, we discuss the results and summarize our conclusions. 

\section{Observations \& Data Reduction}\label{sec:obs&data}
\begin{figure}
\includegraphics[width=\linewidth]{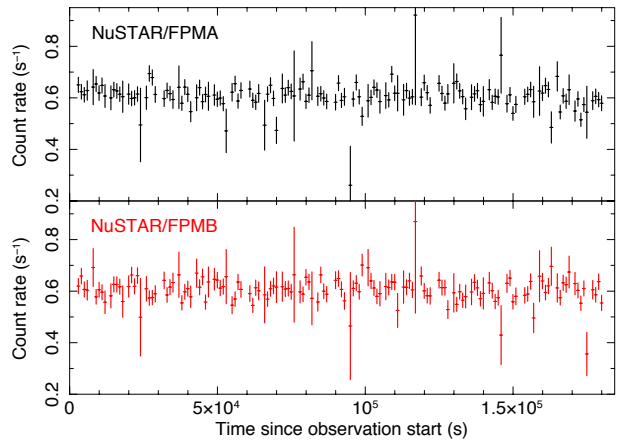}
\caption{The \textit{NuSTAR}/FPMA and FPMB 3-79 keV light curve. Bins of 1000 s are used.
\label{lcurve}}
\end{figure}

\nustar~\citep{harrison2013nustar} observed NGC~7213 starting on 2014 October 5, with a net exposure of 109 ks (Obs. Id. 60001031002). 
The \nustar~data were reduced with the standard pipeline (\textsc{nupipeline}) in the \nustar~Data Analysis Software (\textsc{nustardas}, v1.3.1; part of the \textsc{heasoft} distribution as of version 6.14), using calibration files from \nustar~{\sc caldb} v20141107. Spectra and light curves were extracted from the cleaned event files using the standard tool \textsc{nuproducts} for each of the two hard X-ray telescopes aboard \nustar, which have corresponding focal plane modules A and B (FPMA and FPMB). The spectra from FPMA and FPMB are analysed jointly, but are not combined, allowing for a free cross-calibration constant. The source data were extracted from circular regions (radius 75 arcsec), and background was extracted from a blank area close to the source. Finally, the spectra were binned to have a signal-to-noise ratio (SNR) greater than 3 in each spectral channel, and not to over-sample the instrumental resolution by a factor greater than 2.5. \\ \\
In Fig. \ref{lcurve} we show the \nustar/FPMA and FPMB light curves obtained in the 3--79 keV energy range. 
The total variation in the light curve is of the order of a few per cent. Therefore, in the following we analyse the time-averaged spectrum of the source.
\section{Spectral analysis}\label{sec:analysis}
Spectral analysis and model fitting was carried out with the \textsc{xspec} 12.8 package \citep{arnaud1996}, using the $\chi^2$ minimisation technique throughout. In this work, the errors are quoted at the 90 per cent confidence level, if not stated otherwise.
\subsection{The \nustar~spectrum}\label{subsec:cutoffpl}
\begin{figure}
\includegraphics[width=\linewidth]{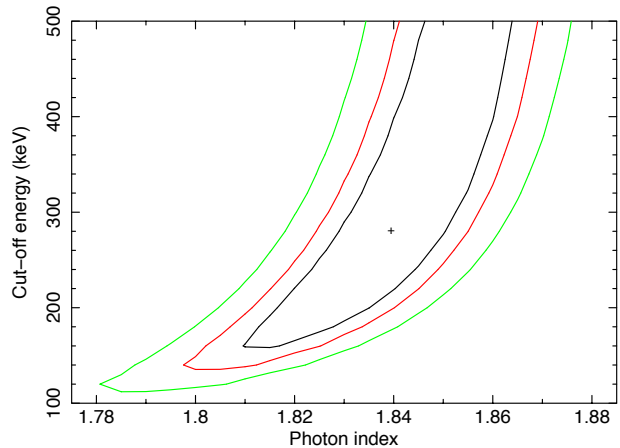}
\caption{Primary continuum cut-off energy vs. photon index contours. Solid green, red and black lines correspond to 99\%, 90\% and 68\% confidence levels, respectively.
\label{gamma_ec}}
\end{figure}

\begin{figure}
\includegraphics[width=\linewidth]{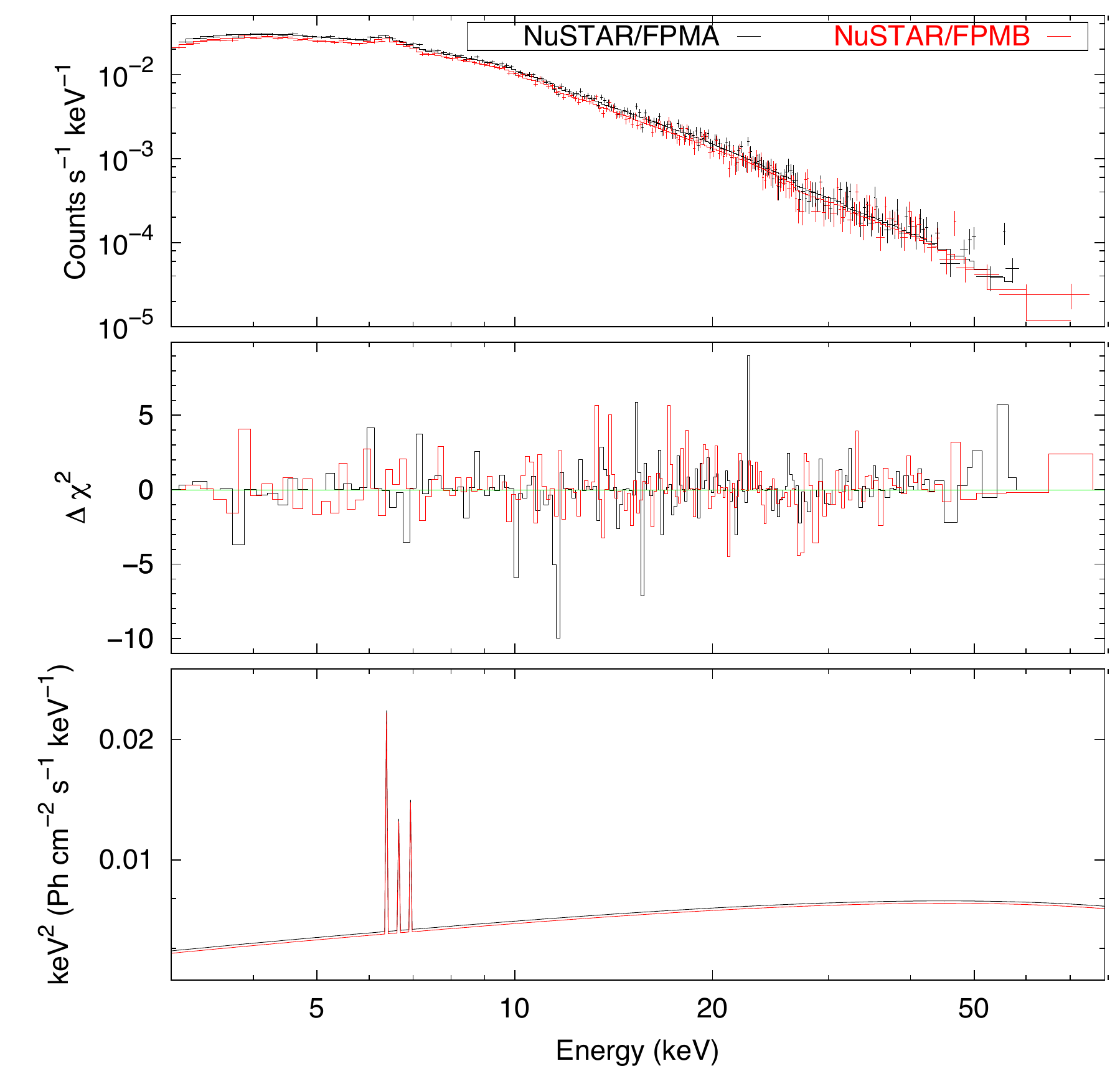}
\caption{The \nustar~spectrum and baseline model (see Table \ref{fit1}). Upper panel: \nustar~data and folded model. Middle panel: contribution to $\chi^2$. Lower panel: best-fitting model $E^2 f(E)$ with the plot of the additive components, namely the cut-off power law and the three \fek~lines.
\label{ldata_chi_eemo}}
\end{figure}

\begin{figure}
\includegraphics[width=\linewidth]{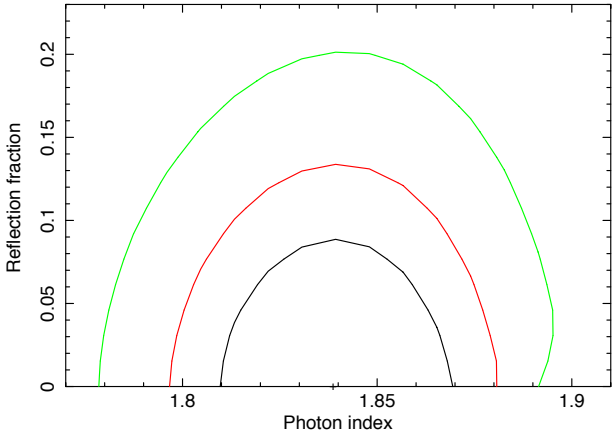}
\caption{\textsc{pexrav} reflection fraction vs. photon index contours. Solid green, red and black lines correspond to 99\%, 90\% and 68\% confidence levels, respectively.
\label{gamma_r}}
\end{figure}

\begin{figure}
\includegraphics[width=\linewidth]{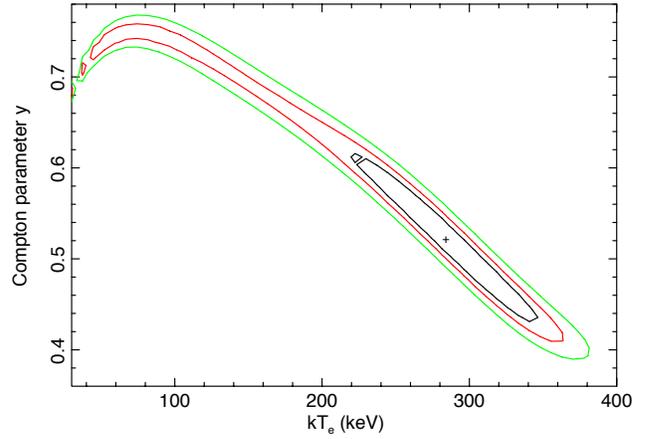}
\caption{\textsc{compps} Compton parameter vs. temperature contours. Solid green, red and black lines correspond to 99\%, 90\% and 68\% confidence levels, respectively.
\label{kt_y}}
\end{figure}

As a first step, we define a phenomenological model, which we describe below. We fit the \nustar~data in the 3--79 keV range, allowing for a free cross-calibration constant between the modules FPMA and FPMB. The two modules are in good agreement, with a cross-calibration factor K$_{A-B}=0.99 \pm 0.01$ fixing the constant for the FPMA data to unity. \\ \\
We modelled the continuum with a cut-off power law, modified by neutral absorption (\textsc{phabs} model in \textsc{xspec}) from Galactic hydrogen with column density $N_\textrm{H} = 2.04 \times 10^{20}$ cm$^{-2}$ \citep{nh_gal}. This simple model yields a poor fit (reduced $\chi^2$ = 556/369), with positive residuals between 6 and 7 keV, which can be attributed to the known Fe complex between 6.4 and 7 keV \citep{bianchi20037213,bianchi72132008}. \citet{bianchi72132008} found three emission lines in the \textit{Chandra}/HEG spectrum, at 6.4, 6.7 and 6.966 keV. These lines are interpreted as emission from neutral Fe, Fe \textsc{xxv} and Fe \textsc{xxvi} respectively. \textit{Chandra}/HEG data provide much higher energy resolution near the Fe complex compared with \textit{NuSTAR}. We thus followed the analysis of \citet{bianchi72132008} and tested for the presence of three narrow Gaussian lines fixing their energies at 6.4, 6.7 and 6.966 keV, leaving the normalizations free to vary. The inclusion of a narrow Gaussian line at 6.4 keV yields a significantly improved fit (reduced $\chi^2$ = 417/368). Adding a second line at 6.7 keV further improves the fit (reduced $\chi^2$ = 388/367), with a probability of chance improvement less than $4 \times 10^{-7}$ according to the $F$-test. The inclusion of a third line at 6.966 keV yields a good fit (reduced $\chi^2$ = 375/367), without prominent residuals and a probability of chance improvement less than $5 \times 10^{-4}$ according to the $F$-test.
	We report the results of this fit in Table \ref{fit1}, and in Fig. \ref{gamma_ec} we show the contour plots of the cut-off energy versus photon index. In Fig. \ref{ldata_chi_eemo} we show the data, residuals and best-fitting model. To further test the presence of a high-energy cut-off, we included the 70-month average \swift/BAT spectrum \citep{bat70}, however the results are essentially unchanged. \\ \\
We tested for the presence of a reflection continuum by including the \textsc{pexrav} model in \textsc{xspec}, which describes neutral Compton reflection of infinite column density in a slab geometry \citep{pexrav}. This model is adequate for reflection off a standard accretion disc, given the high column densities expected for these structures \citep[see, e.g.,][]{sz1994}. We fixed the inclination angle of the reflector to 30 deg, appropriate for a type 1 source \citep[e.g.,][]{nandra1997}. The photon index and cut-off energy of the incident spectrum were tied to those of the primary power law. However, no improvement is found, with only an upper-limit to the reflection fraction $\mathcal{R}$ of 0.13, at 90 per cent confidence level. In Fig. \ref{gamma_r} we show the contour plots of $\mathcal{R}$ versus photon index. This is consistent with the iron lines originating from Compton-thin material ($N_\textrm{H}=10^{22}-10^{23}$ cm$^{-2}$), which does not produce a prominent Compton reflection hump \citep{bianchi20037213}. To test this result further, we replaced the \textsc{pexrav} component and the neutral \fek~line with the \textsc{MYTorus} model, which includes Compton reflection and iron fluorescent lines from a gas torus with an opening angle of 60 deg \citep{mytorus}. The inclination angle of the torus was fixed at 30 deg. The column densities of the scattered and line components were linked and free to vary. The normalizations of the scattered and line components were tied to the normalization of the primary power law, i.e. the standard \textsc{MYTorus} configuration \citep[``coupled'' reprocessor; see, e.g.,][]{mytorus2}. The assumed geometry corresponds to a covering fraction of 0.5. We obtain a good fit (reduced $\chi^2$ = 373/363) and we find a column density of $5.0^{+2.0}_{-1.6}\times 10^{23}$ \sqcm. This result is consistent with the estimate of $\sim 3\times 10^{23}$ \sqcm~for the \fek~line-emitting material reported in \citet{bianchi72132008}\\ \\ 
We next replaced the cut-off power law with the Comptonization model \textsc{compps} \citep{compps} in \textsc{xspec}. \textsc{compps} models the thermal Comptonization emission of a hot plasma cooled by soft photons with a disc black-body distribution. We set a spherical geometry (parameter $\textsc{geom}=-4$ in \textsc{compps}) for the hot plasma and a temperature at the inner disc radius of 10 eV, fitting for the electron temperature $kT_e$ and Compton parameter $y=4 \tau (kT_e / m_e c^2)$. The model yields a good fit (reduced $\chi^2$ = 375/366) with best-fitting parameters $kT_e= 295^{+70}_{-250}$ keV and $y=0.52^{+0.34}_{-0.10}$, which imply an optical depth $\tau = 0.2^{+0.7}_{-0.1}$. In Fig. \ref{kt_y} we show the  $kT_e-y$ contour plots. 
\begin{table}
\begin{center}
\caption{Best-fitting parameters of the baseline model including a primary cut-off power law and three Gaussian lines (line 1 at 6.4 keV, line 2 at 6.7 keV, line 3 at 6.966 keV). \label{fit1}}
\begin{tabular}{ l l } 
\hline \hline \rule{0pt}{2.5ex}   
$\Gamma$ & $1.84\pm 0.03$\\ \rule{0pt}{2.5ex}  
$E_c$ (keV) & $>140$\\ \rule{0pt}{2.5ex}  
$F_{\textrm{3--10 keV}}$ ($10^{-11}$  erg cm$^{-2}$ s$^{-1}$) & $1.3 \pm 0.1$ \\ \rule{0pt}{2.5ex}  
EW$_1$ (eV) & $98^{+30}_{-20}$\\ \rule{0pt}{2.5ex}  
$F_1$ ($10^{-5}$ ph cm$^{-2}$ s$^{-1}$)& $1.7_{-0.4}^{+0.3}$\\ \rule{0pt}{2.5ex}  
EW$_2$ (eV) & $29_{-18}^{+24}$\\ \rule{0pt}{2.5ex}  
$F_2$ ($10^{-5}$ ph cm$^{-2}$ s$^{-1}$)& $0.5_{-0.3}^{+0.5}$\\ \rule{0pt}{2.5ex}  
EW$_3$ (eV) & $42\pm 20$\\  \rule{0pt}{2.5ex}  
$F_3$ ($10^{-5}$ ph cm$^{-2}$ s$^{-1}$) &$0.6 \pm 0.3$ \\  \rule{0pt}{2.5ex}  
$\chi^2$/d.o.f. & 375/366 \\
\hline
 \end{tabular}
\end{center}
\end{table}
\subsection{Time evolution of the Fe K$\alpha$ line}\label{subsec:iron}
\begin{figure}
\includegraphics[width=\linewidth]{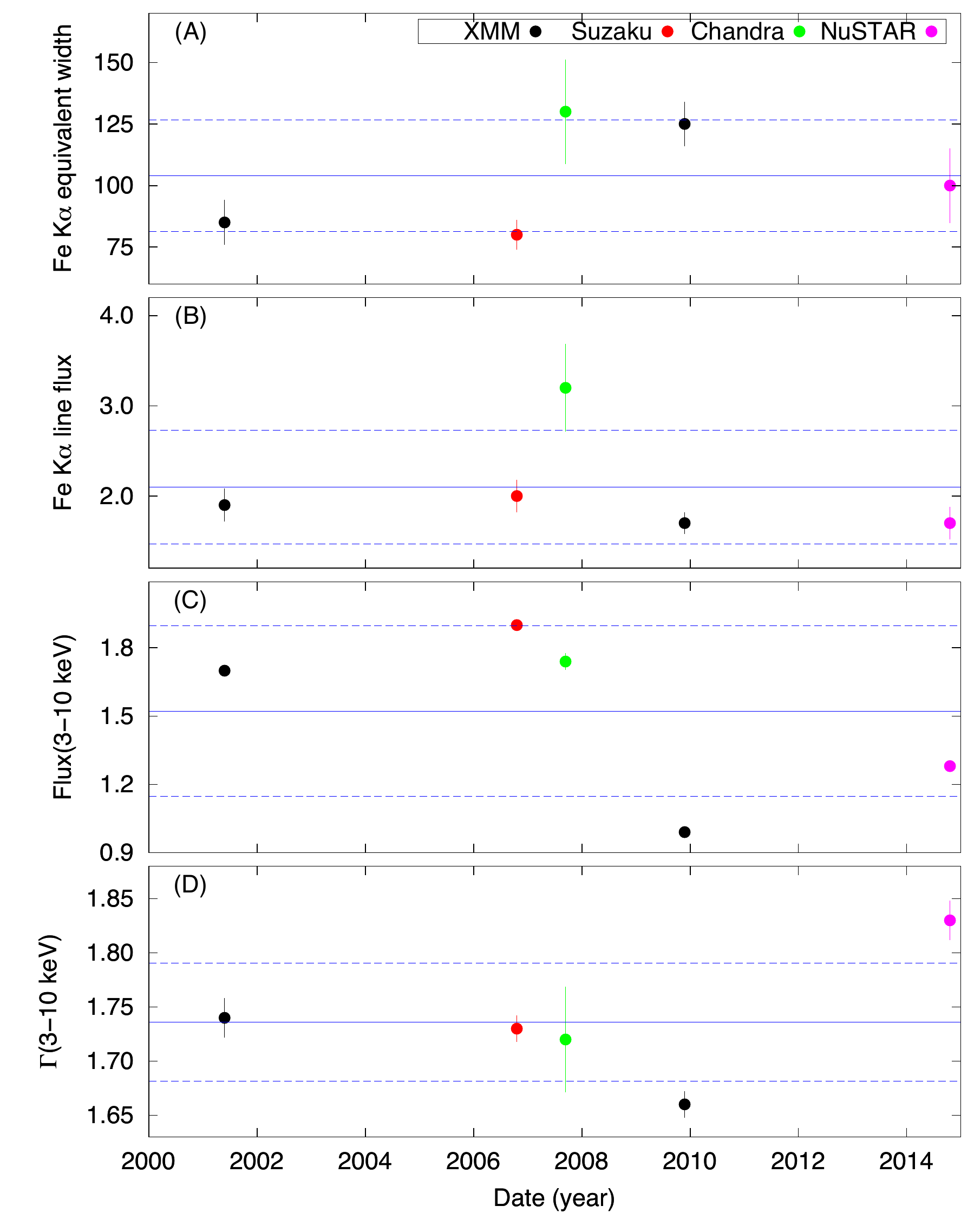}
\caption{Time evolution of \fek~line parameters and 3--10 keV primary flux and photon index. Panel (A): \fek~equivalent width in eV units. Panel (B): \fek~line flux in units of $10^{-5}$ photons \sqcm~s$^{-1}$. Panel (C): continuum 3--10 keV flux in units of $10^{-11}$ erg \sqcm~s$^{-1}$. Panel (D): 3--10 keV photon index. Error bars denote the 1-$\sigma$ uncertainty. The blue solid lines represent the mean value for each parameter, while the blue dashed lines correspond to the standard deviation. 
\label{kalpha}}
\end{figure}

To try to understand the origin of the \fek~lines, we compared our results with past observations of NGC~7213 by re-analysing the archival \xmm~(2001, 2009), \chandra~(2007) and \suzaku~(2006) data. We fitted the 3--10 keV data sets with the baseline model described in Sect. \ref{subsec:cutoffpl}, i.e. including a cut-off power law and three narrow Gaussian lines. The best-fitting parameters are consistent with those reported in \citet{bianchi20037213,bianchi72132008} and \citet{lobban2010}.\\ \\
In Fig. \ref{kalpha} we show the time evolution of the neutral \fek~line EW and flux, and of the 3--10 keV continuum flux and photon index.
The continuum flux variations reach up to a factor of 2, while the line flux varies at the $1 \sigma$ confidence level only during the \chandra/HEG observation of 2007. 
The line EW, on the other hand, exhibits significant variations up to a factor of 1.7. The data show a hint of an anticorrelation between the continuum flux and the line EW, though we cannot formally establish a significant relation. The Spearman's rank correlation coefficient is $-0.4$, but the test returns a $P$ value of 0.42, implying that the null hypothesis is not rejected. Similar results hold for the Fe \textsc{xxv} and Fe \textsc{xxvi} lines, albeit with larger uncertainties on their parameters.
\section{Discussion and conclusions}\label{sec:discussion}
We reported results based on a \nustar~observation of the LLAGN/LINER NGC 7213. We derived constraints on the parameters describing the high-energy (3--79 keV) spectrum, and studied variability over a few year time-scale comparing our results with archival observations of the source.\\ \\
We have been able to constrain the reflection component, finding no evidence for a significant Compton reflection continuum. This is consistent with previous results \citep[e.g.,][]{bianchi20037213,lobban2010,emma2013}. 
We note that the constraints on the reflection component are derived assuming a static X-ray source \citep{pexrav,mytorus}, but it may actually be outflowing \citep{belo1999,malzac2001}. An outflowing, energetically dominant corona can be generated by a geometrically thin, optically thick accretion disc at low accretion rates \citep{merloni&fabian2002}. This scenario is consistent with the lack of a significant optical/UV bump from the disc, if the power output is dominated by the corona. Furthermore, an outflowing X-ray corona could account for the absence of a disc reflection component, since the primary emission would be beamed away from the disc. This has been suggested especially for radio-loud sources having relativistic jets, where the X-ray corona might be the base of the jet itself \citep[see, e.g.,][]{walton2013,lohfink20133c120,ballantyne2014,fabian2014}. Although NGC~7213 is a radio-intermediate source with no clear evidence for a strong jet \citep{blank20057213, bell20117213}, its radio emission is found to be weakly correlated with the X-ray emission \citep{bell20117213}. Therefore, the outflowing corona scenario cannot be ruled out in this source.\\ \\
We confirm the presence of an iron complex between 6.4 and 7 keV \citep{bianchi20037213}, consisting of three narrow K$\alpha$ emission lines from neutral Fe, Fe \textsc{xxv} and Fe \textsc{xxvi}. The lack of a Compton reflection hump above 10 keV indicates that these lines cannot originate from Compton-thick material. Using the \textsc{MYTorus} self-consistent model, we find that the neutral \fek~line may be produced by material with a covering fraction of 0.5 and a column density of $5.0^{+2.0}_{-1.6} \times 10^{23}$ \sqcm. This result is consistent with the iron line complex being produced in the BLR \citep{bianchi72132008}. 
Assuming that the line widths represent a Keplerian velocity, it is possible to estimate the distance of the BLR using the virial theorem, i.e. $R_{\textrm{pc}} \simeq 4.5 \times 10^5 f^{-1} \, M_8 \, (\Delta V_{\textrm{km/s}})^{-2}$, where $R_{\textrm{pc}}$ is the distance in units of parsecs, $f$ is a coefficient that accounts for the unknown geometry and orientation \citep{peterson2004}, $M_8$ is the black hole mass in units of $10^8$ solar masses and $\Delta V_{\textrm{km/s}}$ is the line width in units of km s$^{-1}$. 
\citet{bianchi72132008} report a FWHM of $2400^{+1100}_{-600}$  km s$^{-1}$ for the neutral \fek~line using the \chandra/HEG data, and $2640^{+110}_{-90}$  km s$^{-1}$ for the H$\alpha$ line, using ESO-NTT optical data. Setting $f=4.3$ \citep{grier2013}, the virial theorem yields a BLR distance of $\sim 0.02$ pc. Alternatively, reprocessing may occur in dusty gas, since the presence of dust may enhance the neutral \fek~EW \citep{gohil2015}. This effect is due to a suppression of the reflected continuum, caused by the reduction of backscattering opacity in dusty gas \citep{draine2003}. This hypothesis is supported by the presence of strong silicate emission features in the mid-IR spectrum of NGC~7213 \citep{honig2010,ruschel-dutra2014}. The dust sublimation radius is estimated to be $\sim 0.03$ pc from the optical luminosity \citep{honig2010}. Therefore, if the line originates from dusty gas, the line-emitting region should be more distant than the BLR, as suggested by the constancy of the line flux on a few years time-scale (see below). We also note that the lack of any cold reflection from Compton-thick material is consistent with the models of clumpy tori, which are predicted to disappear in low-luminosity sources \citep{elitzur2006, honig2007}. \\ \\ 
We analysed the neutral \fek~line time evolution over $\sim 15$ years by re-fitting archival \xmm, \suzaku~and \chandra~data of the source. 
According to our estimate, the BLR distance is $\sim 0.02$ pc, or $\sim 20$ light days \citep[see also][]{kaspi2005,balmaverde2014}. The observed X-ray variability time-scale of NGC~7213 is of several days or weeks \citep{emma2012}, in agreement with the lack of variability during the $\sim 100$ ks \nustar~observation. We would thus expect the \fek~line flux to vary in response to the continuum variations, albeit with a time delay due to the light-travel time. However, we only detect a marginal variability of the line flux during the \chandra~observation of 2007. Therefore, a contribution from more distant material cannot be ruled out. 
The \textit{Astro-H} satellite, providing an unprecedented combination of high spectral resolution and effective area at 6--7 keV, will likely enable us to solve the ambiguous origin of the \fek~line. For example, the line might be the superposition of two components, namely a broad component from the BLR and a narrow one produced further away, which \textit{Astro-H} should be able to disentangle \citep{astroh-white}. \\ \\
The primary continuum is well fitted by a power law with a photon index $\Gamma=1.84 \pm0.03$, and an extrapolated 2--10 keV flux $F_{\textrm{2--10}} = (1.6 \pm 0.1) \times 10^{-11}$ erg cm$^{-2}$ s$^{-1}$. The 3--10 keV continuum flux and photon index that we found for the archival observations do not show a clear trend (see Fig. \ref{kalpha}). 
The extrapolated 2--10 keV luminosity is $L_{\textrm{2--10}} = (1.2 \pm 0.1) \times 10^{42}$ erg~s$^{-1}$. Using the 2--10 keV bolometric correction of \citet{marconi2004}, we estimate the bolometric luminosity to be $L_{\textrm{bol}} = (1.3 \pm 0.1) \times 10^{43}$ erg~s$^{-1}$. For a black hole mass of $10^8$ $M_{\odot}$, the Eddington luminosity is $1.2 \times 10^{46}$ erg~s$^{-1}$. These estimates yield an accretion rate of $\sim 0.1$ per cent of the Eddington limit, in rough agreement with the value of  $0.14$ per cent by \citet{emma2012} based on the broad-band spectral energy distribution.
An Eddington ratio $\sim 10^{-3}$ is lower than what is typically found for luminous AGNs, namely $\sim 0.01 - 1$, and it lies within the theoretically predicted RIAF regime \citep{narayan1998,ho2009}.\\ \\
We can only place a lower limit on the presence of a high-energy cut-off, $E_c > 140$ keV, consistent with the lower limit of 350 keV found by \citet{lobban2010} using \suzaku+\textit{Swift}/BAT data, and marginally consistent with the value of $95^{+75}_{-45}$ keV found by \citet{bianchi20037213} using \xmm+\sax~data. 
Replacing the cut-off power law with a Comptonization model and assuming a spherical geometry, we estimate the coronal temperature to be $>$ 40 keV ($kT_e = 295^{+70}_{-250}$ keV) and we measure an optical depth $\tau = 0.2^{+0.7}_{-0.1}$. The lack of an upper limit on the high-energy cut-off does not conflict with the upper limit on the coronal temperature, because a cut-off power law is known to be a rough approximation of Comptonization models \citep[see, e.g.,][]{stern1995}. First, the high-energy turnover of a Comptonization spectrum is much sharper than an exponential cut-off \citep[see, e.g.,][]{zdziarski2003}. Moreover, a Comptonization spectrum is actually a superposition of several orders of Compton scattering spectra. When the optical depth is small, the different scattering orders are separated in energy, thus resulting in a bumpy spectral shape \citep[see, e.g.,][]{compps}. But the optical depth and temperature are inversely related for a given heating/cooling ratio, i.e. the ratio of the power dissipated in the corona to the intercepted soft luminosity \citep[][]{stern1995,hmg1997}. This can produce an upper limit on the coronal temperature even if no exponential cut-off is required \citep[see, e.g.,][]{pop2013mrk509}. \\ \\
The combination of a weak or absent reflected continuum, a weak UV bump, and a low accretion rate suggest that the standard optically thick, geometrically thin accretion disc is truncated in the inner region of the source \citep{starling2005,lobban2010}. A possible explanation is that the nucleus accretes via a RIAF, with the inner edge of any standard disc restricted to large distances, $\sim 10^2$ gravitational radii \citep[see, e.g.,][]{quatar1999}. This is a natural suggestion for a LLAGN, since RIAFs are only expected in sub-Eddington systems and the low radiative efficiency would explain the observed low luminosity. The X-ray emission from a RIAF is likely dominated by thermal Comptonization, in agreement with observations of X-ray binaries in the hard state \citep[see, e.g.,][and references therein]{narayan2005}. The soft seed photons can be synchrotron photons produced in the hot accretion flow itself (synchrotron self-Compton), or thermal photons from the outer thin disc \citep[see, e.g.,][]{yuan&zdziarski,nemmen2014}. As we noted above, the X-ray variability time-scale of NGC~7213 is of the order of days/weeks \citep{emma2012}. Therefore, the Comptonization process is consistent with taking place in an extended region with $kT_e > 40$ keV and $\tau \lesssim 1$, possibly illuminated by the outer thin disc. The relatively small optical depth is a further indication that the corona should be extended and subtend a large solid angle as seen from the disc, in order to scatter a sufficient number of soft photons to produce a substantial X-ray continuum. \\ \\
Finally, our results can be compared with those on other weakly accreting AGNs observed so far by \nustar, namely NGC~5506 \citep[estimated Eddington ratio as low as $ 0.7 \times 10^{-2}$; see][]{matt20155506}\footnote{This value of the Eddington ratio is obtained by assuming the highest estimate of the black hole mass in NGC~5506, namely $\sim 10^8$ $M_{\odot}$. The mass is poorly known, with a lower limit of a few $\times 10^6$ $M_{\odot}$ \citep[see][]{guainazzi5506}, which would yield an Eddington ratio above $0.1$.} and NGC~2110 \citep[estimated Eddington ratio of 0.25--3.7 $\times 10^{-2}$; see][]{marinucci2110}. \citet{matt20155506} found a lower limit on the high-energy cut-off of $\sim 500$ keV for NGC~5506, while \citet{marinucci2110} inferred $E_c > 210$ keV for NGC~2110. In these objects, then, a weak disc emission may be accompanied by a relatively high coronal temperature, in agreement with our present work on NGC~7213. The number of observed sources with good measurements of the high-energy cut-off is too small to establish any statistical correlation with other parameters. However, if the X-ray emission is due to a Comptonizing corona, a high coronal temperature is expected when the disc radiation is weak, because inverse Compton scattering is inefficient in cooling the corona. Further studies on a greater number of sources will be required to confirm this scenario by constraining the physical parameters of the disc/corona system.

\section*{Acknowledgements}
We thank the anonymous referee for his/her helpful comments, which improved the manuscript.
FU thanks Pierre-Olivier Petrucci for useful discussions and comments.
This work is based on observations obtained with the \textit{NuSTAR} mission, a project led by the California Institute of Technology, managed by the Jet Propulsion Laboratory and funded by NASA. This research has made use of data, software and/or web tools obtained from NASA's High Energy Astrophysics Science Archive Research Center (HEASARC), a service of Goddard Space Flight Center and the Smithsonian Astrophysical Observatory. 
FU, GM, SB acknowledge support from the French-Italian International Project of Scientific Collaboration: PICS-INAF project number 181542. FU acknowledges support from CNES and Universit\'e Franco-Italienne (Vinci PhD fellowship). FU, AM, GM acknowledges financial support from the Italian Space Agency under grant ASI/INAF I/037/12/0-011/13. SB acknowledge financial support from the Italian Space Agency under grant ASI-INAF I/037/12/P1. PA acknowledges support from FONDECYT 1140304. FEB acknowledges support from CONICYT-Chile (Basal-CATA PFB-06/2007, FONDECYT 1141218, "EMBIGGEN" Anillo ACT1101), and the Ministry of Economy, Development, and Tourism's Millennium Science Initiative through grant IC120009, awarded to The Millennium Institute of Astrophysics, MAS.
\bibliographystyle{mn2e_mod.bst}
\bibliography{mybib.bib}

\begin{thebibliography}{70}
\providecommand{\natexlab}[1]{#1}

\bibitem[{{Arnaud}(1996)}]{arnaud1996}
{Arnaud} K.~A., 1996, in G.H. {Jacoby}, J.~{Barnes}, eds, Astronomical Data
  Analysis Software and Systems V. Astronomical Society of the Pacific
  Conference Series, Vol. 101, p.~17

\bibitem[{{Ballantyne} et~al.(2014)}]{ballantyne2014}
{Ballantyne} D.~R. et~al., 2014, \apj, 794, 62

\bibitem[{{Balmaverde} \& {Capetti}(2014)}]{balmaverde2014}
{Balmaverde} B., {Capetti} A., 2014, \aap, 563, A119

\bibitem[{{Balokovi{\'c}} et~al.(2015)}]{balokovic2015}
{Balokovi{\'c}} M. et~al., 2015, \apj, 800, 62

\bibitem[{{Baumgartner} et~al.(2013)}]{bat70}
{Baumgartner} W.~H., {Tueller} J., {Markwardt} C.~B., {Skinner} G.~K.,
  {Barthelmy} S., {Mushotzky} R.~F., {Evans} P.~A., {Gehrels} N., 2013, \apjs,
  207, 19

\bibitem[{{Bell} et~al.(2011)}]{bell20117213}
{Bell} M.~E. et~al., 2011, \mnras, 411, 402

\bibitem[{{Beloborodov}(1999)}]{belo1999}
{Beloborodov} A.~M., 1999, \apjl, 510, L123

\bibitem[{{Bianchi} et~al.(2003){Bianchi}, {Matt}, {Balestra} \&
  {Perola}}]{bianchi20037213}
{Bianchi} S., {Matt} G., {Balestra} I., {Perola} G.~C., 2003, \aap, 407, L21

\bibitem[{{Bianchi} et~al.(2008)}]{bianchi72132008}
{Bianchi} S., {La Franca} F., {Matt} G., {Guainazzi} M., {Jimenez Bail{\'o}n}
  E., {Longinotti} A.~L., {Nicastro} F., {Pentericci} L., 2008, \mnras, 389,
  L52

\bibitem[{{Blank} et~al.(2005){Blank}, {Harnett} \& {Jones}}]{blank20057213}
{Blank} D.~L., {Harnett} J.~I., {Jones} P.~A., 2005, \mnras, 356, 734

\bibitem[{{Brenneman} et~al.(2014)}]{IC4329A_Brenneman}
{Brenneman} L.~W. et~al., 2014, \apj, 788, 61

\bibitem[{{Dickey} \& {Lockman}(1990)}]{nh_gal}
{Dickey} J.~M., {Lockman} F.~J., 1990, \araa, 28, 215

\bibitem[{{Draine}(2003)}]{draine2003}
{Draine} B.~T., 2003, \apj, 598, 1026

\bibitem[{{Elitzur} \& {Shlosman}(2006)}]{elitzur2006}
{Elitzur} M., {Shlosman} I., 2006, \apjl, 648, L101

\bibitem[{{Emmanoulopoulos} et~al.(2012){Emmanoulopoulos}, {Papadakis},
  {McHardy}, {Ar{\'e}valo}, {Calvelo} \& {Uttley}}]{emma2012}
{Emmanoulopoulos} D., {Papadakis} I.~E., {McHardy} I.~M., {Ar{\'e}valo} P.,
  {Calvelo} D.~E., {Uttley} P., 2012, \mnras, 424, 1327

\bibitem[{{Emmanoulopoulos} et~al.(2013){Emmanoulopoulos}, {Papadakis},
  {Nicastro} \& {McHardy}}]{emma2013}
{Emmanoulopoulos} D., {Papadakis} I.~E., {Nicastro} F., {McHardy} I.~M., 2013,
  \mnras, 429, 3439

\bibitem[{{Fabian} et~al.(2014){Fabian}, {Parker}, {Wilkins}, {Miller}, {Kara},
  {Reynolds} \& {Dauser}}]{fabian2014}
{Fabian} A.~C., {Parker} M.~L., {Wilkins} D.~R., {Miller} J.~M., {Kara} E.,
  {Reynolds} C.~S., {Dauser} T., 2014, \mnras, 439, 2307

\bibitem[{{Filippenko} \& {Halpern}(1984)}]{filippenko1984}
{Filippenko} A.~V., {Halpern} J.~P., 1984, \apj, 285, 458

\bibitem[{{Gohil} \& {Ballantyne}(2015)}]{gohil2015}
{Gohil} R., {Ballantyne} D.~R., 2015, \mnras, 449, 1449

\bibitem[{{Grier} et~al.(2013)}]{grier2013}
{Grier} C.~J. et~al., 2013, \apj, 773, 90

\bibitem[{{Gu} \& {Cao}(2009)}]{gu&cao2009}
{Gu} M., {Cao} X., 2009, \mnras, 399, 349

\bibitem[{{Haardt} \& {Maraschi}(1991)}]{haardt&maraschi1991}
{Haardt} F., {Maraschi} L., 1991, \apjl, 380, L51

\bibitem[{{Haardt} et~al.(1994){Haardt}, {Maraschi} \& {Ghisellini}}]{hmg1994}
{Haardt} F., {Maraschi} L., {Ghisellini} G., 1994, \apjl, 432, L95

\bibitem[{{Haardt} et~al.(1997){Haardt}, {Maraschi} \& {Ghisellini}}]{hmg1997}
{Haardt} F., {Maraschi} L., {Ghisellini} G., 1997, \apj, 476, 620

\bibitem[{{Harrison} et~al.(2013)}]{harrison2013nustar}
{Harrison} F.~A. et~al., 2013, \apj, 770, 103

\bibitem[{{Ho}(2008)}]{ho2008review}
{Ho} L.~C., 2008, \araa, 46, 475

\bibitem[{{Ho}(2009)}]{ho2009}
{Ho} L.~C., 2009, \apj, 699, 626

\bibitem[{{H{\"o}nig} \& {Beckert}(2007)}]{honig2007}
{H{\"o}nig} S.~F., {Beckert} T., 2007, \mnras, 380, 1172

\bibitem[{{H{\"o}nig} et~al.(2010)}]{honig2010}
{H{\"o}nig} S.~F., {Kishimoto} M., {Gandhi} P., {Smette} A., {Asmus} D.,
  {Duschl} W., {Polletta} M., {Weigelt} G., 2010, \aap, 515, A23

\bibitem[{{Kaspi} et~al.(2005){Kaspi}, {Maoz}, {Netzer}, {Peterson},
  {Vestergaard} \& {Jannuzi}}]{kaspi2005}
{Kaspi} S., {Maoz} D., {Netzer} H., {Peterson} B.~M., {Vestergaard} M.,
  {Jannuzi} B.~T., 2005, \apj, 629, 61

\bibitem[{{Kollmeier} et~al.(2006)}]{kollmeier2006}
{Kollmeier} J.~A. et~al., 2006, \apj, 648, 128

\bibitem[{{Lobban} et~al.(2010){Lobban}, {Reeves}, {Porquet}, {Braito},
  {Markowitz}, {Miller} \& {Turner}}]{lobban2010}
{Lobban} A.~P., {Reeves} J.~N., {Porquet} D., {Braito} V., {Markowitz} A.,
  {Miller} L., {Turner} T.~J., 2010, \mnras, 408, 551

\bibitem[{{Lohfink} et~al.(2013)}]{lohfink20133c120}
{Lohfink} A.~M. et~al., 2013, \apj, 772, 83

\bibitem[{{Magdziarz} \& {Zdziarski}(1995)}]{pexrav}
{Magdziarz} P., {Zdziarski} A.~A., 1995, \mnras, 273, 837

\bibitem[{Malizia et~al.(2014)Malizia, Molina, Bassani, Stephen, Bazzano,
  Ubertini \& Bird}]{2041-8205-782-2-L25}
Malizia A., Molina M., Bassani L., Stephen J.~B., Bazzano A., Ubertini P., Bird
  A.~J., 2014, \apj, 782, L25

\bibitem[{{Malzac} et~al.(2001){Malzac}, {Beloborodov} \&
  {Poutanen}}]{malzac2001}
{Malzac} J., {Beloborodov} A.~M., {Poutanen} J., 2001, \mnras, 326, 417

\bibitem[{{Maoz}(2007)}]{maoz2007}
{Maoz} D., 2007, \mnras, 377, 1696

\bibitem[{{Marconi} et~al.(2004){Marconi}, {Risaliti}, {Gilli}, {Hunt},
  {Maiolino} \& {Salvati}}]{marconi2004}
{Marconi} A., {Risaliti} G., {Gilli} R., {Hunt} L.~K., {Maiolino} R., {Salvati}
  M., 2004, \mnras, 351, 169

\bibitem[{{Marinucci} et~al.(2014)}]{marinucci2014swift}
{Marinucci} A. et~al., 2014, \mnras, 440, 2347

\bibitem[{{Marinucci} et~al.(2015)}]{marinucci2110}
{Marinucci} A. et~al., 2015, \mnras, 447, 160

\bibitem[{{Matt} et~al.(2015)}]{matt20155506}
{Matt} G. et~al., 2015, \mnras, 447, 3029

\bibitem[{{Murphy} \& {Yaqoob}(2009)}]{mytorus}
{Murphy} K.~D., {Yaqoob} T., 2009, \mnras, 397, 1549

\bibitem[{{Nandra} et~al.(1997){Nandra}, {George}, {Mushotzky}, {Turner} \&
  {Yaqoob}}]{nandra1997}
{Nandra} K., {George} I.~M., {Mushotzky} R.~F., {Turner} T.~J., {Yaqoob} T.,
  1997, \apj, 477, 602

\bibitem[{{Narayan}(2005)}]{narayan2005}
{Narayan} R., 2005, \apss, 300, 177

\bibitem[{{Narayan} \& {Yi}(1994)}]{adaf}
{Narayan} R., {Yi} I., 1994, \apjl, 428, L13

\bibitem[{{Nemmen} et~al.(2014){Nemmen}, {Storchi-Bergmann} \&
  {Eracleous}}]{nemmen2014}
{Nemmen} R.~S., {Storchi-Bergmann} T., {Eracleous} M., 2014, \mnras, 438, 2804

\bibitem[{{Panessa} et~al.(2006)}]{panessa2006}
{Panessa} F., {Bassani} L., {Cappi} M., {Dadina} M., {Barcons} X., {Carrera}
  F.~J., {Ho} L.~C., {Iwasawa} K., 2006, \aap, 455, 173

\bibitem[{{Pereira-Santaella} et~al.(2010){Pereira-Santaella},
  {Diamond-Stanic}, {Alonso-Herrero} \& {Rieke}}]{D7213}
{Pereira-Santaella} M., {Diamond-Stanic} A.~M., {Alonso-Herrero} A., {Rieke}
  G.~H., 2010, \apj, 725, 2270

\bibitem[{{Perola} et~al.(2002)}]{perola2002}
{Perola} G.~C., {Matt} G., {Cappi} M., {Fiore} F., {Guainazzi} M., {Maraschi}
  L., {Petrucci} P.~O., {Piro} L., 2002, \aap, 389, 802

\bibitem[{{Peterson} et~al.(2004)}]{peterson2004}
{Peterson} B.~M. et~al., 2004, \apj, 613, 682

\bibitem[{{Petrucci} et~al.(2013)}]{pop2013mrk509}
{Petrucci} P.~O. et~al., 2013, \aap, 549, A73

\bibitem[{{Phillips}(1979)}]{phillips1979}
{Phillips} M.~M., 1979, \apjl, 227, L121

\bibitem[{{Poutanen} \& {Svensson}(1996)}]{compps}
{Poutanen} J., {Svensson} R., 1996, \apj, 470, 249

\bibitem[{{Quataert} et~al.(1999){Quataert}, {Di Matteo}, {Narayan} \&
  {Ho}}]{quatar1999}
{Quataert} E., {Di Matteo} T., {Narayan} R., {Ho} L.~C., 1999, \apjl, 525, L89

\bibitem[{{Reynolds} et~al.(2014)}]{astroh-white}
{Reynolds} C. et~al., 2014, ArXiv e-prints

\bibitem[{{Ruschel-Dutra} et~al.(2014){Ruschel-Dutra}, {Pastoriza}, {Riffel},
  {Sales} \& {Winge}}]{ruschel-dutra2014}
{Ruschel-Dutra} D., {Pastoriza} M., {Riffel} R., {Sales} D.~A., {Winge} C.,
  2014, \mnras, 438, 3434

\bibitem[{{Shakura} \& {Sunyaev}(1973)}]{ss1973}
{Shakura} N.~I., {Sunyaev} R.~A., 1973, \aap, 24, 337

\bibitem[{{Sobolewska} \& {Papadakis}(2009)}]{sobolewska&papadakis}
{Sobolewska} M.~A., {Papadakis} I.~E., 2009, \mnras, 399, 1597

\bibitem[{{Starling} et~al.(2005){Starling}, {Page}, {Branduardi-Raymont},
  {Breeveld}, {Soria} \& {Wu}}]{starling2005}
{Starling} R.~L.~C., {Page} M.~J., {Branduardi-Raymont} G., {Breeveld} A.~A.,
  {Soria} R., {Wu} K., 2005, \mnras, 356, 727

\bibitem[{{Stern} et~al.(1995){Stern}, {Poutanen}, {Svensson}, {Sikora} \&
  {Begelman}}]{stern1995}
{Stern} B.~E., {Poutanen} J., {Svensson} R., {Sikora} M., {Begelman} M.~C.,
  1995, \apjl, 449, L13

\bibitem[{{Svensson} \& {Zdziarski}(1994)}]{sz1994}
{Svensson} R., {Zdziarski} A.~A., 1994, \apj, 436, 599

\bibitem[{{Ursini} et~al.(2015)}]{5548}
{Ursini} F. et~al., 2015, \aap, 577, A38

\bibitem[{{Walton} et~al.(2013){Walton}, {Nardini}, {Fabian}, {Gallo} \&
  {Reis}}]{walton2013}
{Walton} D.~J., {Nardini} E., {Fabian} A.~C., {Gallo} L.~C., {Reis} R.~C.,
  2013, \mnras, 428, 2901

\bibitem[{{Woo} \& {Urry}(2002)}]{woo&urry2002}
{Woo} J.~H., {Urry} C.~M., 2002, \apj, 579, 530

\bibitem[{{Wu} et~al.(1983){Wu}, {Boggess} \& {Gull}}]{wu1983}
{Wu} C.~C., {Boggess} A., {Gull} T.~R., 1983, \apj, 266, 28

\bibitem[{{Wu} \& {Gu}(2008)}]{wu&gu2008}
{Wu} Q., {Gu} M., 2008, \apj, 682, 212

\bibitem[{{Yaqoob}(2012)}]{mytorus2}
{Yaqoob} T., 2012, \mnras, 423, 3360

\bibitem[{{Yu} et~al.(2011){Yu}, {Yuan} \& {Ho}}]{yu2011}
{Yu} Z., {Yuan} F., {Ho} L.~C., 2011, \apj, 726, 87

\bibitem[{{Yuan} \& {Zdziarski}(2004)}]{yuan&zdziarski}
{Yuan} F., {Zdziarski} A.~A., 2004, \mnras, 354, 953

\bibitem[{{Zdziarski} et~al.(2003){Zdziarski}, {Lubi{\'n}ski}, {Gilfanov} \&
  {Revnivtsev}}]{zdziarski2003}
{Zdziarski} A.~A., {Lubi{\'n}ski} P., {Gilfanov} M., {Revnivtsev} M., 2003,
  \mnras, 342, 355

\end{thebibliography}
\end{document}